\documentclass{pssbmac}

\usepackage[english]{babel} 

\usepackage[utf8]{inputenc} 

\usepackage{listings}
\usepackage{xcolor}
\usepackage[colorlinks,linkcolor=blue]{hyperref}

\usepackage[T1]{fontenc}
\usepackage{float}
\usepackage{graphics}
\usepackage{graphicx}
\usepackage{epsfig}
\usepackage{indentfirst}
\usepackage{amsmath, amsfonts, amssymb, amsthm, mathtools}
\usepackage{url}
\usepackage{csquotes}

\usepackage[backend=biber, style=ieee, maxnames=50]{biblatex}
\DeclareTextFontCommand{\emph}{\boldmath\bfseries}
\DefineBibliographyStrings{english}{mathesis = {Master dissertation}}

\begin{document}


\title{Implementation of the Multichannel Filtered Reference Least Mean Square (McFxLMS) Algorithm with an Arbitrary Number of Channels by Using MATLAB}

\author{
    {\large Boxiang Wang}\thanks{BOXIANG001@e.ntu.edu.sg} 
}
\criartitulo


\begin{abstract}
Multichannel filtered reference least mean square (McFxLMS) algorithms are widely utilized in adaptive multichannel active noise control (MCANC) applications. As a critical and high-computationally efficient adaptive critical algorithm, it also typically works as a benchmark for comparative studies of the new algorithms proposed by peers and researchers. However, up to now, there are few open-source codes for the FxLMS algorithm, especially for large-count channels. Therefore, this work provides a MATLAB code for the McFxLMS algorithm, which can be used for the arbitrary number of channels system. The code is available on \href{https://github.com/ShiDongyuan/Multichannel_FxLMS_Matlab.git}{GitHub} and \href{https://www.mathworks.com/matlabcentral/fileexchange/158731-multichannel-filtered-x-least-mean-square-mcfxlms}{MathWorks}. 

\end{abstract}

\section{Introduction}

Active Noise Control (ANC) is an advanced technique used to reduce unwanted sound by introducing controlled anti-noise that effectively cancels the original noise. This process, known as destructive interference, is a sophisticated application of the principle of superposition in wave physics. Due to its efficiency in reducing low-frequency noise and compact size, ANC is widely utilized in many applications sensitive to noise interferences~\cite{lam2021ten,shi2023active1}, such as headphones~\cite{shen2023implementations,shen2022multi,shen2022hybrid,shen2022adaptive,shen2021alternative,shen2021wireless}, automobiles, and windows~\cite{lam2018active,lam2020active, lam2020active1,lam2023anti,shi2017algorithmsC,hasegawa2018window,shi2017understanding,lai2023robust,shi2016open}. Compared to the single-channel structure~\cite{shi2016systolic,shi2019two,wen2020convergence}, the Multichannel Active Noise Control (MCANC) system owns more freedom of controllable degree and, hence, is more effective in dealing with complicated acoustic environments~\cite{shi2019practical}. Furthermore, to assist MCANC in coping with the dynamic noise and varying acoustic environments, adaptive algorithms, such as the multichannel filtered reference least mean square (McFxLMS) algorithm, have been developed to use in noise cancellation. Despite the development of numerous sophisticated algorithms~\cite{luo2023delayless,luo2023gfanc,luo2023deep,ji2023practical,luo2022hybrid,shi2022selective,shi2023transferable,shi2021comb,shi2023frequency,luo2023performance,shi2019optimal,shi2021optimal,shi2021fast} and methods~\cite{shi2020feedforward,ji2023computation,shi2020feedforward,shi2018novel,lai2023mov,lai2023real,shen2023momentum} to improve the efficiency of ANC systems, McFxLMS plays a crucial role in active control and serves as the benchmark for these new algorithms. In addition, its derivative algorithms~\cite{luo2022implementation,shi2017multiple,shi2021block,shi2019analysis,shi2020active,shi2020multichannel,shi2023multichannel,shi2023computation2,shi2024behindN} are flourishing to address the practical obstacles of actual ANC applications. Therefore, this study employs the widely-used simulation tool, MATLAB, to realize the McFxLMS algorithm and make its code accessible to the public. This code can be readily employed to implement the co-located and fully connected structures. 

The document provides a detailed explanation of a MATLAB program that simulates feedforward multichannel active noise control (ANC) using the filtered-x least mean square (FxLMS) algorithm. Specifically, the user is free to choose the number of reference microphones, the number of secondary sources, and the number of error microphones. In addition, the program offers the option of using the GPU to accelerate the computation.

\section{Brief Theoretical Introduction on Multichannel Filtered Reference Least Mean Square Algorithm}
Active noise control (ANC) is a mechanism used to address low-frequency noise issues based on the principle of acoustic wave superposition~\cite{dongyuan2020algorithms}. The ANC system artificially generates an anti-noise wave that has the same amplitude but the reverse phase of the noise wave, which interferes with the disturbance destructively.

In general, the ANC system can be classified as feedforward structure and feedback structure. The feedforward structure implements the reference microphone and the error microphone to generate anti-noise that can dynamically match with the variation of the primary noise, which allows it to deal with many noise types. Moreover, the ANC system also can be referred to as single-channel ANC~\cite{shi2016systolic} or multichannel ANC based on the number of secondary sources used. Compared to single-channel ANC, multichannel ANC is implemented to gain a larger quiet zone through multiple secondary sources and error microphones. 

The FxLMS is among the most practical adaptive algorithms proposed to compensate for the influence of the secondary path in an ANC system. Figure 1 shows the block diagram of the multichannel FxLMS algorithm, which has $J$ reference microphones, $K$ secondary sources, and $M$ error microphones. The control filter matrix is given by

\begin{equation}
    \mathbf w(n) = \begin{bmatrix}
        \mathbf{w}_{11}^T(n) & \mathbf{w}_{12}^T(n) & \cdots & \mathbf{w}_{1J}^T(n) \\
        \mathbf{w}_{21}^T(n) & \mathbf{w}_{22}^T(n) & \cdots & \mathbf{w}_{2J}^T(n) \\
        \vdots & \vdots & \ddots & \vdots \\
        \mathbf{w}_{K1}^T(n) & \mathbf{w}_{K2}^T(n) & \cdots & \mathbf{w}_{KJ}^T(n)
    \end{bmatrix} \in \mathbb{R}^{K \times JN},
\end{equation}
where \( \mathbf{w}_{kj}(n) = [ w_{kj,0}(n), w_{kj,1}(n), \ldots, w_{kj,N-1}(n) ]^T \in \mathbb{R}^{JN \times 1} \) is the control filter from the $j$th input to the $k$th output, and $N$ denotes the length of the control filter. The reference vector is given by
\begin{equation}
    \mathbf x(n) = \left[ \mathbf{x}_1^T(n), \mathbf{x}_2^T(n), \ldots, \mathbf{x}_J^T(n) \right]^T \in \mathbb{R}^{JN \times 1},
\end{equation}
where \(\mathbf{x}_j = \left[ x_j(n), x_j(n - 1), \ldots, x_j(n - N + 1) \right]^T \in \mathbb{R}^{N \times 1}\). The output of the control filter is given by 
\begin{equation}
\mathbf y(n) = \mathbf w(n) \mathbf x(n) \in \mathbb{R}^{K \times 1}.
\end{equation}

\begin{figure}[H]
\centering
\includegraphics[width=.7\textwidth]{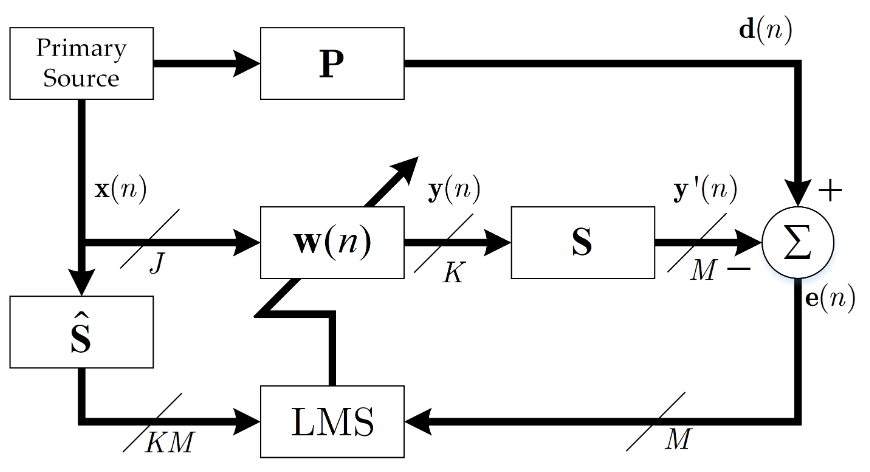}
\caption{ {\small Block diagram of the multichannel FxLMS algorithm.}}
\label{figure 1}
\end{figure}

The anti-noise vector at the error microphones, as illustrated in Figure 1, is given by
\begin{equation}
\mathbf y'(n) = \mathbf s * \mathbf y(n) \in \mathbb{R}^{M \times 1},
\end{equation}
where $\mathbf{s}$ represents the impulse response matrix of secondary paths
\begin{equation}
    \mathbf{s} = \begin{bmatrix}
        \mathbf{s}_{11}(n) & \mathbf{s}_{12}(n) & \cdots & \mathbf{s}_{1K}(n) \\
        \mathbf{s}_{21}(n) & \mathbf{s}_{22}(n) & \cdots & \mathbf{s}_{2K}(n) \\
        \vdots & \vdots & \ddots & \vdots \\
        \mathbf{s}_{M1}(n) & \mathbf{s}_{M2}(n) & \cdots & \mathbf{s}_{MK}(n)
    \end{bmatrix} \in \mathbb{R}^{M \times K}
    .
\end{equation}

The error signal vector $\mathbf{e}(n)$ measured by $M$ error microphones can be expressed as
\begin{equation}
    \mathbf e(n) = \mathbf d(n) - \mathbf y'(n) \in \mathbb{R}^{M \times 1},
\end{equation}
where \(\mathbf d(n) = \left[ d_1(n), d_2(n), \ldots, d_M(n) \right]^T \in \mathbb{R}^{M \times 1} \) stands for the disturbance vector, with $d_m(n)$ denoting the disturbance at the $m$th $(m = 1, 2, · · · , M )$ error microphone. The cost function of the adaptive filter is defined as
\begin{equation}
    J(n) = \sum_{m=1}^{M} e_m^2(n).
\end{equation}

We replace the impulse response $s_{mk}(n)$ with its estimate $\hat{\mathbf{s}}_{mk}(n)$ to give
\begin{equation}
    x_{jkm}'(n) = \hat{s}_{mk}(n) * x_j(n).
\end{equation}

The control filter weight vector $\mathbf{w}_{kj}(n)$ is updated in the direction of the negative gradient with step size $\mu$ based on the minimization of the estimated cost function $J(n)$:
\begin{equation}
    \mathbf{w}_{kj}(n + 1) = \mathbf{w}_{kj}(n) + \mu \sum_{m=1}^{M} e_m(n) \mathbf{x}'_{jkm}(n),
\end{equation}
where the filtered reference vector is
\begin{equation}
\mathbf{x}'_{jkm}(n) = \left[ x'_{jkm}(n), x'_{jkm}(n - 1), \ldots, x'_{jkm}(n - N + 1) \right]^T.    
\end{equation}

\section{Code Explanation}
This MATLAB program implements the simulation of a feedforward multichannel FxLMS ANC system, with the number of channels that can be specified arbitrarily by the user. The program contains three MATLAB files: \textcolor{blue}{$Multichannel\_FxLMS.m$} and \textcolor{blue}{$Control\_filter.m$} constitute the program of the McFxLMS algorithm, while \textcolor{red}{$Four\_MCANC.m$} is the main program to test the McFxLMS algorithm.  

\subsection{Control\_filter}
The control filter is an essential element of the ANC system, responsible for processing the reference signal to generate the control signal. 

Therefore, \textcolor{blue}{$Control\_filter.m$} defines a class called $\textbf{Control\_filter}$, which specifies the properties and functions of the control filter for a multichannel ANC system and is an important component that will be used in the \textcolor{blue}{$Multichannel\_FxLMS.m$} class.

\subsubsection{Properties of Control\_filter}
The properties of $\textbf{Control\_filter}$ are shown below.
\begin{lstlisting}[language=Matlab]
properties
    Sec_Matrix %--The coefficients of the secondary path (Ls x E_num).
    umW        %--The step size of the algorithm.
    Len        %--The length of the control filter.
    Ls         %--The length of the secondary path.
    Wc         %--The control filter coefficients. 
    Fd         %--The buffer of the control filter (Len x E_num).
    E_num      %--The number of the error microphones.
    Xd         %--The delay line of the reference.
    Xf         %--The delay line of the filtered reference.
    Yd         %--The delay line of the control filter output.
end
\end{lstlisting}

In this code snippet, $Sec\_Matrix$ is used to store the coefficients of the secondary path and has a dimension of $Ls$ by $E\_num$; $Ls$ and $E\_num$ denote the length of the secondary path impulse response and the number of the error microphones, respectively. $Len$ represents the length of the control filter; $Wc$ is the vector of the control filter. $Xd$, $Xf$, and $Yd$ stands for the delay line of the input data. 

\newpage
\subsubsection{Function 1: Control\_filter}
This function is used to initialize the class instance, which sets the properties of the control filter.  In particular, it initializes some properties based on whether $Sec\_Matrix$ is a gpuArray to support GPU-accelerated computation.
\begin{lstlisting}[language=Matlab]
function obj=Control_filter(Sec_Matrix, Len, umW)
    obj.Len        = Len        ;
    obj.umW        = umW        ;
    csize          = size(Sec_Matrix);
    obj.Ls         = csize(1)   ;
    obj.E_num      = csize(2)   ;
    obj.Sec_Matrix = Sec_Matrix ;
    if isa(Sec_Matrix,'gpuArray')
        obj.Wc = gpuArray(zeros(Len,1));
        obj.Xd = gpuArray(zeros(Len,1));
        obj.Xf = gpuArray(zeros(obj.Ls,1));
        obj.Yd = gpuArray(zeros(obj.Ls,1));
        obj.Fd = gpuArray(zeros(Len,obj.E_num));
    else
        obj.Wc = zeros(Len,1);
        obj.Xd = zeros(Len,1);
        obj.Xf = zeros(obj.Ls,1);
        obj.Yd = zeros(obj.Ls,1);
        obj.Fd = zeros(Len,obj.E_num);
    end
end
\end{lstlisting}
\newpage

\subsubsection{Function 2: Generator\_antinoise}
This function is the key function of $Control\_filter$ that generates the anti-noise signal. It receives the input signal $xin$ and the error signal $Er$, then updates the filter coefficients $Wc$ according to the FxLMS algorithm to generate the anti-noise signal $y\_anti$. This function also updates various delay lines and the buffer of the control filter.
\begin{lstlisting}[language=Matlab]
function [y_anti, obj] = Generator_antinoise(obj, xin, Er)
    umW1= obj.umW ;
    Xd1 = obj.Xd  ;
    Xf1 = obj.Xf  ;
    Wc1 = obj.Wc  ;
    Fd1 = obj.Fd  ;
    Yd1 = obj.Yd  ;
    Sec_Matrix1 = obj.Sec_Matrix;
    %---------------------------->>>----------------------<<<
    Xd1  = [xin;Xd1(1:end-1)];
    Xf1  = [xin;Xf1(1:end-1)];
    Wc1  = Wc1 - umW1 * Fd1 * Er;
    y_o  = Wc1'*Xd1 ; 
    Yd1  = [y_o;Yd1(1:end-1)];
    y_anti = Sec_Matrix1' * Yd1;
    fd  = Sec_Matrix1' * Xf1 ;
    Fd1  = [fd'; Fd1(1:end-1,:)];
    %---------------------------->>>----------------------<<<
    obj.Xd = Xd1;
    obj.Xf = Xf1;
    obj.Wc = Wc1;
    obj.Fd = Fd1;
    obj.Yd = Yd1;
end
\end{lstlisting}

\newpage
\subsection{Multichannel\_FxLMS}
\textcolor{blue}{$Multichannel\_FxLMS.m$} defines a class called $Multichannel\_FxLMS$, which specifies the properties and functions of the multichannel ANC system.
\subsubsection{Properties of Multichannel\_FxLMS}
The properties of $Multichannel\_FxLMS$ are shown below.
\begin{lstlisting}[language=Matlab]
properties
    controller %--The controller of the multichannel system.
    Cunit      %--The number of the control units.
    R_num      %--The number of reference microphones.
    E_num      %--The number of error microphones.
    S_num      %--The number of secondary sources.
    lenc       %--The length of the control filter.
end
\end{lstlisting}

\subsubsection{Function 1: Multichannel\_FxLMS}
This function is used to initialize the class instance, which sets the properties of the multichannel ANC system. Specifically, this function supports two types of multichannel ANC systems: fully-connection ANC and collocated ANC based on the number of control units $Cunit$. In addition, information such as system structure (fully-connection or collocated) and hardware usage (CPU or GPU) is printed out.
\newpage
\begin{lstlisting}[language=Matlab]
function obj = Multichannel_FxLMS(Wc,Sec_estimate,umW)
    w_size = size(Wc); 
    len   = w_size(1); 
    Cunit = w_size(2); 
    if length(w_size)== 2
        R_num = 1;
    else
        R_num = w_size(3); 
    end
    obj.lenc  = len  ;
    obj.Cunit = Cunit;
    obj.R_num = R_num;
    
    s_size = size(Sec_estimate);
    E_num  = s_size(2); 
    if length(s_size)==2
        S_num  = 1;
    else
        S_num  = s_size(3);
    end
    obj.E_num = E_num ;
    obj.S_num = S_num ;
    obj.controller = repmat(Control_filter(squeeze(Sec_estimate(:,:,1)),len, umW),Cunit,R_num);
    if Cunit ~= 1  % Full structure 
        for jj = 1:R_num 
            for kk = 1:Cunit
                obj.controller(kk,jj) = Control_filter(squeeze(Sec_estimate(:,:,kk)),len, umW);                   
            end
        end
        disx3 = sprintf('Structure: Fully-connection feedforward ANC.');
    else         % Collocated structure 
        for jj = 1:R_num 
            obj.controller(1,jj) = Control_filter(squeeze(Sec_estimate(:,:,jj)) ,len, umW);
        end
        disx3 = sprintf('Structure: Collocated feedforward ANC.');
    end 
    fprintf('<<--------------------------------------------------->>\n');
    fprintf('The multichannel FxLMS has been sucessfuly created.\n');
    disp(disx3);
    if isa(Wc,'gpuArray')
        fprintf('Hardware usage: GPU.\n');
    else
        fprintf('Hardware usage: CPU.\n');
    end
    fprintf('Dimension: %d x %d x %d \n', R_num, S_num, E_num);
    fprintf('<<--------------------------------------------------->>\n');
end
\end{lstlisting}

\subsubsection{Function 2: FxLMS\_canceller}
This function is the key function of $Multichannel\_FxLMS$ that implements the FxLMS algorithms to achieve multichannel ANC. It receives the reference signal $Re$ and the disturbance signal $Disturbance$ as inputs then calls the $Generator\_antinoise$ function of $control\_filter$ moment by moment to generate the anti-noise signal. Finally, the anti-noise signal is added to the disturbance signal $Disturbance$ to obtain the error signal $Er$.

\begin{lstlisting}[language=Matlab]
function [Er, obj] = FxLMS_canceller(obj, Re, Disturbance)
    % Re: the reference matrix [microphone x signal length]
    % Disturabnce: the disturbance matrix [microphone x signal length]
    data_size = size(Re)    ;
    len       = data_size(2); 
    if isa(Re,'gpuoArray')
        E = gpuArray(zeros(obj.E_num,len+1));
    else
        E = zeros(obj.E_num,len+1); 
    end
    %% Filtering processing
    tic
    fprintf('<<-------------------START-----------------------------\n');
    for nn = 1: len 
        if obj.Cunit ~=1
            for jj = 1:obj.R_num
                for kk = 1:obj.Cunit
                    [Y,obj.controller(kk,jj)] = Generator_antinoise(obj.controller(kk,jj),
                    Re(jj,nn),E(:,nn));
                    E(:,nn+1) = E(:,nn+1) + Y ; 
                end
            end
        else 
            for jj = 1:obj.R_num
                [Y,obj.controller(1,jj)] = Generator_antinoise(obj.controller(1,jj),
                Re(jj,nn),E(:,nn));
                 E(:,nn+1) = E(:,nn+1) + Y ; 
            end
        end
        E(:,nn+1) = Disturbance(:,nn) + E(:,nn+1);
    end
    Er = gather(E(:,2:end));
    toc
    fprintf('---------------------END----------------------------->>\n');
end
\end{lstlisting}

\newpage
\subsubsection{Function 3: Get\_coefficients}
This function is used to obtain the coefficients of the control filters in the multichannel ANC system.
\begin{lstlisting}[language=Matlab]
function Wc = Get_coefficients(obj)
    if obj.Cunit ~=1
        Wc = zeros(obj.lenc,obj.Cunit,obj.R_num);
        for jj = 1:obj.R_num
            for kk = 1:obj.Cunit
                contorler = obj.controller(kk,jj);
                Wc(:,kk,jj)=contorler.Wc;
            end
        end
    else
        Wc = zeros(obj.lenc,obj.R_num);
        for jj = 1:obj.R_num
            contorler = obj.controller(1,jj);
            Wc(:,jj)=contorler.Wc;
       end
   end
end
\end{lstlisting}

\section{Testing code: Four channel active noise cancellation}
\begin{figure}
    \centering
    \includegraphics[width=11cm]{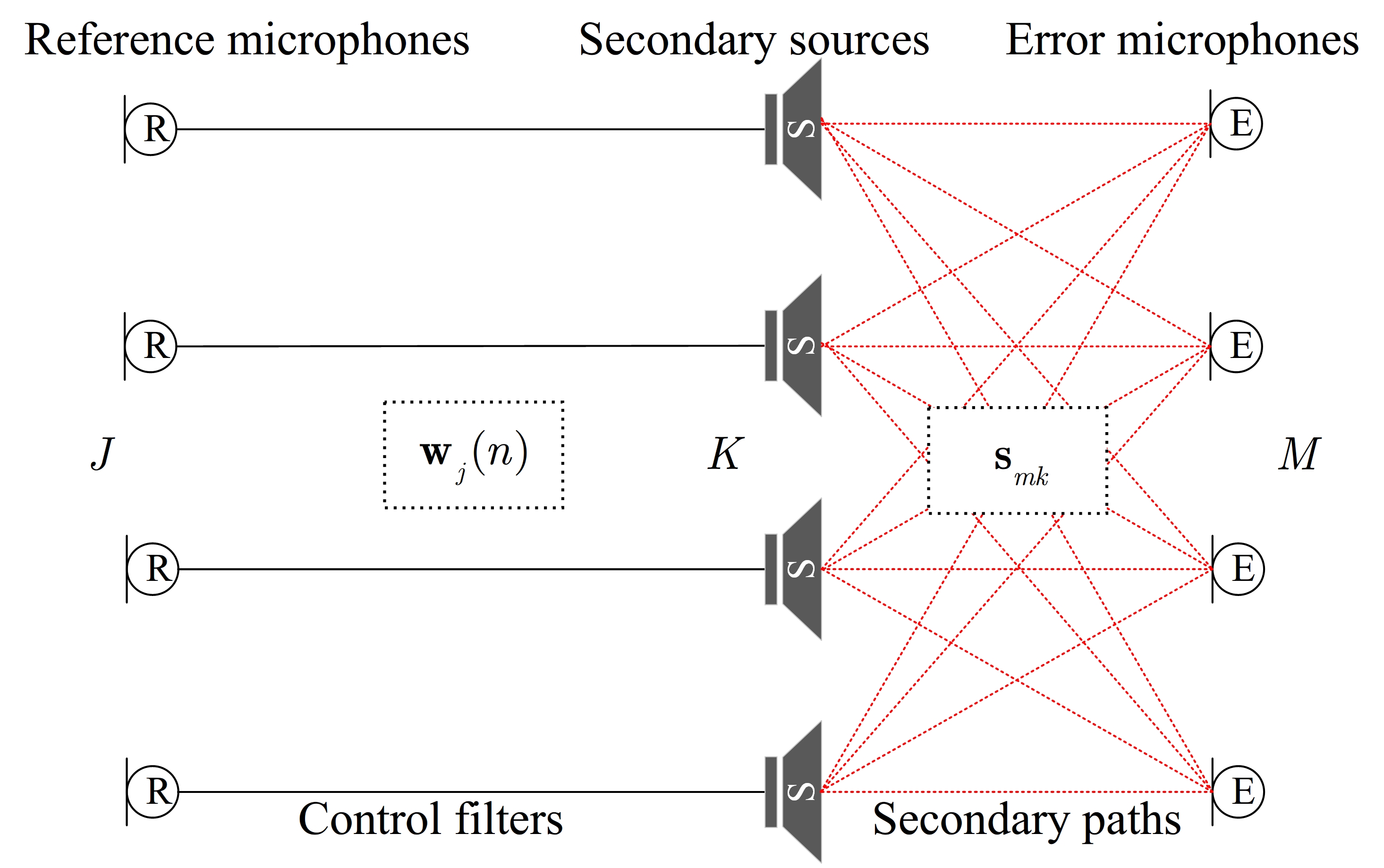}
    \caption{Block diagram of the 4-channel collocated active noise control system.}
    \label{fig:enter-label}
\end{figure}

The $Four\_MCANC.m$ carries out a simulation on a collocated 4x4x4 ANC system, as shown in Figure~\ref{fig:enter-label}. In the simulation, $Multichannel\_FxLMS.m$ and $Four\_MCANC.m$ are used to reduce the disturbance. 
\subsection{Loading primary and secondary paths}
The primary path is the path the noise takes from the noise source to the error microphone and the secondary path is the path the anti-noise takes from the secondary source to the error microphone. Here, the primary paths and the secondary paths are loaded from files and the length of both the primary paths and the secondary paths are 256.
\begin{lstlisting}[language=Matlab]
Pri = zeros(256,4); % Primary path
for nn=1:4
    a = sprintf('path\\P%d.mat',nn);
    b = load(a);
    c = sprintf('b.P%d',nn);
    d = eval(c)   ;
    Pri(:,nn) = d ;
end
Sec = zeros(256,4,4); % Secondary path
for ss = 1:4 % Number of secondary sources
    for mm = 1:4 % Number of error microphones
        a = sprintf('path\\S%d%d.mat',ss,mm);
        b = load(a);
        c = sprintf('b.S%d%d',ss,mm);
        d = eval(c)   ;
        Sec(:,mm,ss)=d;
    end
end
\end{lstlisting}

\subsection{Generate reference and disturbance signals}
The following commands generate a broadband noise of 400-800 Hz as reference signals by filtering white noise through a bandpass filter, and generate disturbance signals by filtering the reference signal through the primary path. The sampling frequency of the system is set to 16 kHz. 

\begin{lstlisting}[language=Matlab]
fs =  16000; % Sampling frequency
t  =  40; % Sampling time
T  = 0:1/fs:t ;
len= length(T); % Signal length
Re = randn(len,1); % Generate white noise
bf = fir1(512,[0.05 0.1]); % Bandpass filter with passband 400-800Hz
Re = filter(bf,1,Re); % Filtering the white noise
Re1 = [Re';Re';Re';Re']; % Reference [reference microphone x signal length] 
Dir = zeros(4,len); % Disturbance [reference microphone x signal length]
for jj = 1:4
    Dir(jj,:) = (filter(Pri(:,jj),1,Re1(jj,:)))';
end
Red = Re1;
\end{lstlisting}

\newpage
\subsubsection{Noise cancellation based on the McFxLMS algorithm}
The following commands simulate a collocated 4x4x4 ANC system with the length of control filters is 512 and the stepsize of the FxLMS algorithm is 0.00001.
\begin{lstlisting}[language=Matlab]
Wc  = zeros(512,1,4); % Implement a 4-by-4-by-4 channel FxLMS.
muW = 0.00001; % Stepsize of the FxLMS algorithm
a = Multichannel_FxLMS(Wc,Sec,muW);
[E,a]= a.FxLMS_cannceller(Red,Dir);
\end{lstlisting}

\subsection{Drawing the figure of error signals}
Drawing the error signals of the four error microphones.
\begin{lstlisting}[language=Matlab]
subplot(2,2,1)
plot(E(1,:));
xlim([0 650000])
xlabel('Time')
ylabel('Amplitude')
title('Error Microphone 1')
grid on ;
subplot(2,2,2)
plot(E(2,:));
xlim([0 650000])
xlabel('Time')
ylabel('Amplitude')
title('Error Microphone 2')
grid on ;
subplot(2,2,3)
plot(E(3,:));
xlim([0 650000])
xlabel('Time')
ylabel('Amplitude')
title('Error Microphone 3')
grid on ;
subplot(2,2,4)
plot(E(4,:));
xlim([0 650000])
xlabel('Time')
ylabel('Amplitude')
title('Error Microphone 4')
grid on ;
\end{lstlisting}

\begin{figure}[H]
\centering
\includegraphics[width=.7\textwidth]{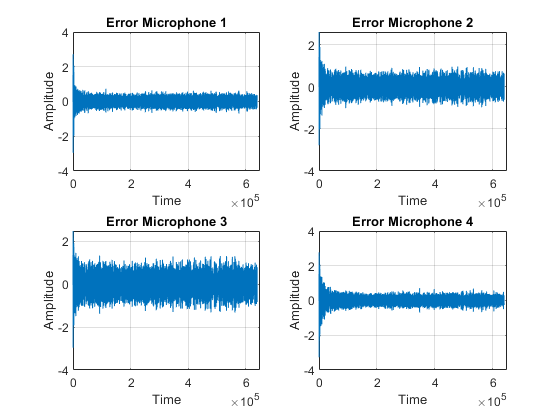}
\caption{ {\small Simulated error signals of the MCANC system using the FxLMS algorithm.}}
\label{figure 2}
\end{figure}

\subsection{Drawing the figure of control filter coefficients}
Drawing the control filter coefficients of the four control filters.
\begin{lstlisting}[language=Matlab]
W1=a.Get_coefficients();
subplot(2,2,1)
plot(W1(:,1));
xlabel('Index')
ylabel('Amplitude')
title('Control Filter 1')
grid on ;
subplot(2,2,2)
plot(W1(:,2));
xlabel('Index')
ylabel('Amplitude')
title('Control Filter 2')
grid on ;
subplot(2,2,3)
plot(W1(:,3));
xlabel('Index')
ylabel('Amplitude')
title('Control Filter 3')
grid on ;
subplot(2,2,4)
plot(W1(:,4));
xlabel('Index')
ylabel('Amplitude')
title('Control Filter 4')
grid on ;
\end{lstlisting}

\begin{figure}[H]
\centering
\includegraphics[width=.7\textwidth]{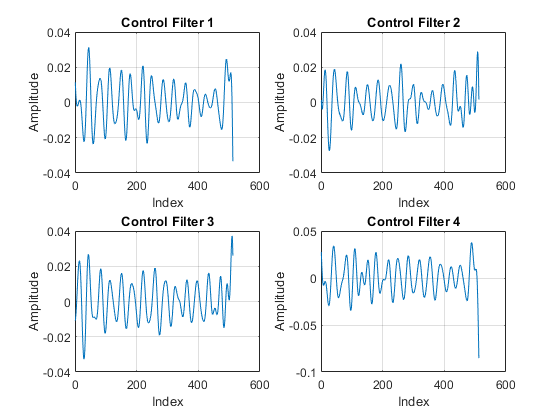}
\caption{ {\small Simulated control filters coefficients of the MCANC system using the FxLMS algorithm.}}
\label{figure 3}
\end{figure}

\section{Conclusion}
The article offers an elaborate elucidation of a MATLAB program that emulates feedforward multichannel active noise control (ANC) by utilizing the filtered-x least mean square (FxLMS) technique. More precisely, the user has the freedom to select the quantity of reference microphones, secondary sources, and error microphones. Furthermore, the program provides the choice to utilize the GPU to enhance the computational efficacy.

\printbibliography


\end{document}